\documentclass[twocolumn,showpacs,aps,prb,superscriptaddress,floatfix]{revtex4}
\usepackage{graphicx}
\usepackage{dcolumn}
\usepackage{bm}

\begin{document}

\preprint{}

\title{On the nature and anisotropy of $\bm{Q}\rm \neq 0$ correlations in Tb$_{2}$Ti$_{2}$O$_{7}$ under the application of magnetic fields} 

\newcommand{\tto}{Tb$_{2}$Ti$_{2}$O$_{7}$}
\newcommand{\hto}{Ho$_{2}$Ti$_{2}$O$_{7}$}
\newcommand{\dto}{Dy$_{2}$Ti$_{2}$O$_{7}$}
\author{J.P.C. Ruff}
\affiliation{Department of Physics and Astronomy, McMaster University,
Hamilton, Ontario, L8S 4M1, Canada}
\author{B.D. Gaulin} 
\affiliation{Department of Physics and Astronomy, McMaster University,
Hamilton, Ontario, L8S 4M1, Canada}
\affiliation{Brockhouse Institute for Materials Research, McMaster University,
Hamilton, Ontario, L8S 4M1, Canada}
\affiliation{Canadian Institute for Advanced Research, 180 Dundas Street West, Toronto, Ontario, Canada M5G 1Z8}
\author{K.C. Rule} 
\affiliation{Helmholtz Zentrum Berlin f\"ur Materialien und Energie, D-14109 Berlin, Germany}
\author{J.S. Gardner} 
\affiliation{NIST Center for Neutron Research, National Institute of Standards and Technology, Gaithersburg, MD 20899-6102, USA}
\affiliation{Department of Physics, Indiana University, 2401 Milo B. Sampson Lane, Bloomington, IN 47408-1398, USA}

\begin{abstract} 

We report an analysis of neutron diffraction from single crystals of the spin-liquid pyrochlore Tb$_{2}$Ti$_{2}$O$_{7}$ under the application of magnetic fields along the crystallographic $[110]$ direction.  Such a perturbation has been shown to destroy the spin liquid ground state and induce long-range order, although the nature of the ordered state was not immediately determined.  Recently, it has been proposed that the ordered state is characterized by spin-ice-like correlations, evincing an emergent ferromagnetic tendency in this material despite the large negative Curie-Weiss constant.  Here, we argue instead that the ordered state is dominated by $\bm{Q}\rm \neq 0$ correlations that emerge either from strong antiferromagnetism or magnetoelastic distortion of the crystal.  In contrast to previous reports, we observe no evidence for re-entrant behaviour in the high field limit.  Extreme sensitivity of the ordered state to the alignment of the applied field is suggested to account for these discrepancies.

\end{abstract} 

\maketitle 

As originally conceived by Anderson,\cite{anderson} a spin liquid is a dynamically disordered quantum antiferromagnetic (AFM) phase wherein strong magnetic correlations exist, but do not extend beyond near-neighbour distances.  This behaviour is dependent on geometrical frustration in the crystal lattice, where the local connectivity of the magnetic ions prohibits the simultaneous satisfaction of all near-neighbour interactions.\cite{diep}  This frustration is commonly realized by a combination of triangular or tetrahedral architectures and AFM interactions.  Ferromagnetic interactions can also be frustrated in the presence of strong single-ion anisotropy, as occurs in the spin ices.\cite{GBrev}

The search for materials that exhibit the ideal spin liquid physics envisioned by Anderson has occupied experimentalists for decades, with mixed results.\cite{ramnat}  Typically, these efforts have focussed on the highly frustrated kagome and pyrochlore lattices, in two and three dimensions respectively.  In general, it has been difficult to identify a material with spin 1/2 moments that crystallizes into a perfect kagome or perfect pyrochlore lattice, where only near-neighbour couplings are relevant.  However, a host of interesting materials with slightly more complicated interactions have been identified, and they exhibit physics that is intriguing in its own right.\cite{diep}

{\tto} belongs to the family of rare-earth-metal titanate pyrochlores, which have garnered recent attention due to the exotic frustration-driven physics they manifest.\cite{MJGrev}  {\tto} is isostructural to the canonical spin ices {\hto} and {\dto},\cite{GBrev} with magnetic rare-earth ions occupying the pyrochlore lattice (Fig. 1).  In contrast to these frustrated ferromagnets, {\tto} presents a significantly reduced single-ion anisotropy ($\Delta \sim 18$ K) as well as a negative Curie-Weiss constant of $\theta_{CW}\sim-19$ K.\cite{Gingras2000}  Interest in {\tto} was ignited when it was observed that dynamic short-range magnetic correlations appear on cooling through $T \sim \theta_{CW}$, but that long range magnetic order is absent down to the lowest measurable temperatures ($\sim 20$ mK).\cite{Gardner99}  These are the key experimental signatures of a spin liquid ground state.  However, {\tto} is expected to be governed by a more complicated Hamiltonian than those generally considered to model spin liquids, since the magnetic moments of the Tb$^{3+}$ ions are large and strongly influenced by spin-orbit coupling and the crystal electric field.  One would therefore expect the moments in {\tto} to be reduced from the free ion value, and to be oriented along local $<$111$>$ directions that join the centers of the corner-sharing tetrahedra in the lattice.\cite{Gingras2000}  This has been experimentally confirmed,\cite{mirebeauice,mirebeaunew} and yet a puzzle remains:  naively, one would expect single-ion anisotropy to lift the degeneracy of the spin liquid ground state, and drive the system to a non-colinear ordered N\'eel state.\cite{Gingras2000,enjalran}  The manner in which frustration is reintroduced is an open topic, although recent work has shown that quantum fluctuations out of the Ising-like ground state doublet can renormalize the effective interactions to be ferromagnetic, leading to a fluctuating quantum spin ice state at low temperatures.\cite{hamid}  There also exists experimental evidence pointing to magnetoelastic effects at low temperatures,\cite{ruffxray,aleksandrov} which suggests that spin-lattice coupling is an important ingredient of the ground state.

\begin{figure} 
\centering 
\includegraphics[width=8.5cm]{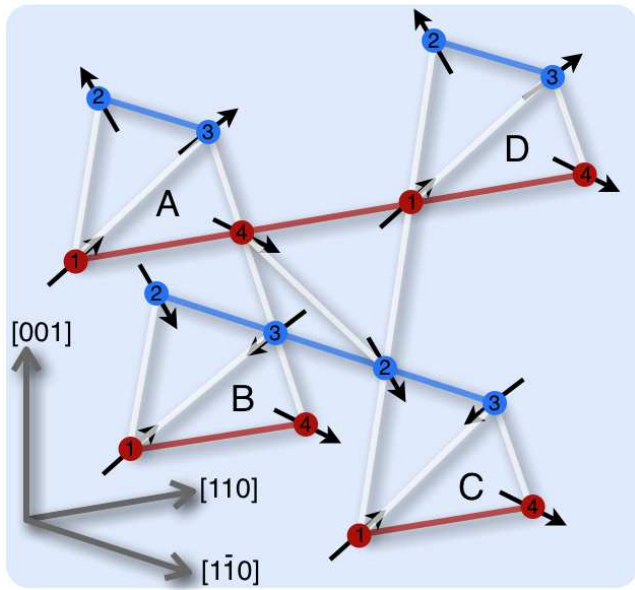}
\caption {(Colour online) The pyrochlore lattice.  The four tetrahedra in the unit cell are labeled by (A,B,C,D) and the four magnetic sublattices are labeled by (1,2,3,4).  Chains running along the $[110]$ applied magnetic field direction are known as $\alpha$-chains, and are shown in red.  Perpendicular $\beta$-chains are shown in blue.  A classical expectation for AFM Ising spins in high field is shown.}
\label{fig:1}
\end{figure}

\begin{table}
\caption{Phase factors in the 16 spin unit cell}
\label{table:1}
\begin{ruledtabular}
\begin{tabular}{c c c c c}
Tetrahedron & Sublattice & $(002)$ & $(1\bar13)$ & $(1\bar12)$ \\ \hline
A & 1 $(\alpha)$ & 1 & 1 & 1 \\
 & 2 $(\beta)$ & -1 & -1 & \it{i}\rm \\
 & 3 $(\beta)$ & -1 & 1 & -\it{i}\rm \\
 & 4 $(\alpha)$ & 1 & 1 & 1 \\ \hline
B & 1 $(\alpha)$ & 1 & 1 & -1 \\
 & 2 $(\beta)$ & -1 & -1 & -\it{i}\rm \\
 & 3 $(\beta)$ & -1 & 1 & \it{i}\rm \\
 & 4 $(\alpha)$ & 1 & 1 & -1 \\ \hline
C & 1 $(\alpha)$ & 1 & 1 & -1 \\
 & 2 $(\beta)$ & -1 & -1 & -\it{i}\rm \\
 & 3 $(\beta)$ & -1 & 1 & \it{i}\rm \\
 & 4 $(\alpha)$ & 1 & 1 & -1 \\ \hline
D & 1 $(\alpha)$ & 1 & 1 & 1 \\
 & 2 $(\beta)$ & -1 & -1 & \it{i}\rm \\
 & 3 $(\beta)$ & -1 & 1 & -\it{i}\rm \\
 & 4 $(\alpha)$ & 1 & 1 & 1 \\ 

\end{tabular}
\end{ruledtabular}
\end{table}

Regardless of the exact origin of the exotic ground state in {\tto}, it is now well known experimentally that long range order can be induced by the application of magnetic fields\cite{rule,mirebeauice} or pressure.\cite{mirebeaunature}  It is not unreasonable to expect that an understanding of these perturbation-induced ordered phases could unravel some of the mysteries of the unperturbed ground state.  Therefore, a careful characterization of the ordered states is crucial.  The original discovery of magnetic field induced order \cite{rule} did not collect enough magnetic Bragg reflections to facilitate a full magnetic structure refinement, although it did identify the high-field ordering wavevector as $\bm{Q} = (2\pi/a)(1\bar12)$, where a \textit{sharp, intense} Bragg reflection was observed at a $\bm{Q}\rm \neq 0$ position.  A weaker, diffuse feature centered at $(1\bar12)$ was also observed, which forms a halo around the sharp peak.  This is evident in Fig. 1c in Ref. \onlinecite{rule}.  Subsequent studies involving many more magnetic reflections returned qualitatively different results,\cite{mirebeauice} finding instead that the dominant scattering was at the $\bm{Q}\rm = 0$ positions.  This was interpreted as evidence for the predicted spin ice like character of {\tto}.  These measurements found weaker evidence for the $\bm{Q}\rm \neq 0$ phase, with a reduced volume fraction and a transition temperature that was suppressed at high fields\cite{mirebeauice} relative to the original work.  This phase was refined as a polarization of spins with Ising axes that project along the magnetic field (generally referred to as $\alpha$-chains), and AFM ordering of spins with Ising axes perpendicular to the magnetic field (known as $\beta$-chains).  A depiction of this phase with a full volume fraction, which is an AFM analogue to the $\bm Q \rm = X$ phase in the spin ices,\cite{fennell,ruff05} is illustrated in Fig. 1.  Assuming comparable sample quality, the main difference between the original measurements of Rule \it{et al.}\rm \cite{rule} and the subsequent measurements of Cao \it{et al.}\rm \cite{mirebeauice} is that the original experiment was performed with the magnetic field aligned along $[110]$ to within an accuracy of $\pm 0.5^{\circ}$, within the mosaic spread of the single crystal sample, while the subsequent experiment was performed with the magnetic field misset from $[110]$ by $\sim 5^{\circ}$ according to the authors.\cite{mirebeauice}  A magnetic field misalignment of this order has been shown to qualitatively change the ground state in the sister spin-ice compounds, both experimentally \cite{clancy} and theoretically.\cite{melko}  Therefore, it is interesting to investigate the discrepancies between these two measurements in further detail.

Here we report a careful analysis of Bragg reflections collected using time-of-flight neutron scattering, in the case where the magnetic field is nominally perfectly aligned (i.e. to within the mosaic spread of the crystal).  We present parametric studies of the nuclear allowed $(1\bar13)$ peak and the superlattice $(002)$ and $(1\bar12)$ peaks. Although insufficient for a brute-force magnetic structure refinement, the relative intensities of these reflections strongly constrain the nature of the ordered state.  Measurements were performed using the Disk Chopper Spectrometer at the NIST Center for Neutron Research, with cold (5\AA) incident neutrons.\cite{DCS}  This time-of-flight instrument is immune to higher-order wavelength contamination problems systematic to diffractometers employing a crystal monochromator.  Bragg intensities were collected as a function of temperature and magnetic field using an 11.5 Tesla superconducting magnet system with a dilution insert, and normalized to incoherent vanadium scattering to correct for relative detector efficiencies and cryostat dark angles.  The results are shown in Fig. 2, where we plot the results of the raw measurement in 2a and 2b, and a useful rescaling of the data in 2c and 2d.  This rescaling is performed as follows: the zero-field component is subtracted out, and intensities are scaled to incoherent vanadium scattering to correct for efficiency.  Then, we multiply by the magnitude of the momentum transfer, take the square root, and divide by the magnetic form factor calculated in the dipole approximation.  The resulting quantity is directly proportional to the total scattering length in the unit cell which gives rise to the Bragg peak, and therefore also proportional to the related ordered magnetic moment transverse to $\bm Q \rm$.  It is clear from both the raw and rescaled intensities that the dominant scattering in the field-induced ordered phase occurs at the superlattice $(1\bar12)$ position.  It is important to note that the intensity of this Bragg peak continues to increase up to the highest applied magnetic field (9 Tesla), in contrast to the re-entrant behaviour recently reported to set in above $\sim$ 5 Tesla.\cite{mirebeauice}  The two preceding points illustrate a clear and qualitative difference between the results presented here, and those reported by Cao \it{et al.}\rm\cite{mirebeauice}  They report weak scattering at the $(1\bar12)$ position that is further suppressed for large applied magnetic fields; we report exactly the opposite.

\begin{figure}
\centering  
\includegraphics[width=8.5cm]{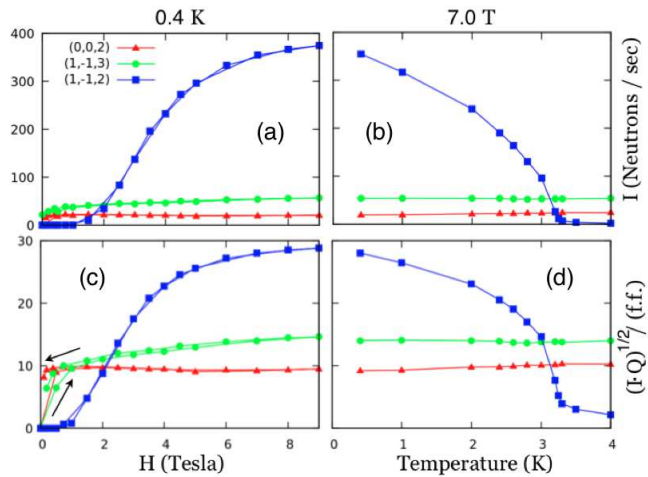}
\caption {(Colour online) Temperature and applied magnetic field dependence of Bragg scattering at three reciprocal space positions.  2a and 2b show integrated intensities measured by rocking the crystal through the Bragg angle, as a function of magnetic field and temperature.  2c and 2d show a rescaling of this data as described in the text. The low-field hysteresis shown in (c) has been previously discussed.\cite{rule}  The remnant intensity at high temperature at the $(1\bar12)$ position is due to the diffuse halo discussed in the text.}
\label{fig:2}
\end{figure}

It is possible to assign qualitative significance to the intensities of each of the measured Bragg reflections, if we assume that the scattering is magnetic in origin.  Table 1 shows the phase factor $\it{e}^{i\bm Q r\rm}$ calculated for each of the three reflections, and for each spin in the cubic nuclear unit cell.  The magnetic ion sites are labeled to coincide with Fig. 1, where each of the four unit cell tetrahedra (A,B,C,D) contain four spins with different easy-axes (1,2,3,4).  It becomes immediately clear that ferromagnetic ordering or field-induced polarization will not contribute to the scattering at $(1\bar12)$, and that AFM correlations between (A,D) tetrahedra and (B,C) tetrahedra will.  Further, an AFM moment on the $\beta$-chains does not interfere with an AFM moment on the $\alpha$-chains - the two contributions are decoupled since the former contributes to the imaginary part of the cross-section, while the latter contributes to the real part.  It is also evident that a net polarization of $\alpha$-chains will contribute to scattering at $(1\bar13)$ but a net polarization of $\beta$-chains will not, and that $(002)$ scattering is a measure of the vector difference in net polarization between the $\alpha$- and $\beta$-chains.  Therefore, despite the fact that this subset of reflections is insufficient to fully refine the magnetic structure, we can confidently comment on the size of the AFM ordered moment, the difference in polarization between the two sets of chains, and the degree to which the $\alpha$-chains are polarized along the applied magnetic field.  By normalizing to the nuclear scattering at $(1\bar13)$ measured and high and low temperature in zero field, we can extract a lower limit on the sizes of these moments which is constrained only by an incomplete understanding of extinction effects in our crystal, which we neccessarily ignore.

Let us begin by considering the AFM analogue of the $\bm Q \rm = X$ spin ice phase\cite{fennell,ruff05} (QXAFM phase), which is shown in Fig. 1.  This is a classical ground state for an Ising AFM in high field, where field perpendicular spins (on $\beta$-chains) order into AFM chains, with a phase difference of $\pi$ between the (A,D) tetrahedra and the (B,C) tetrahedra, while $\alpha$-chain spins maximize their polarization along the field.  Although it is possible to construct a $\bm Q \rm = 0$ state in this way as well, where (A,D) $\beta$-chain spins are not flipped relative to (B,C) $\beta$-chain spins, such a configuration would not generate scattering at $(1\bar12)$.  The classical expectation would be that the states described above are degenerate for nearest neghbour interactions, so the selection of $\bm{Q}\rm \neq 0$ order is indicative of the importance of long-range interactions.  A ground state resembling QXAFM was refined as a minority phase in the work of Cao \it{et al.}\rm,\cite{mirebeauice} but here we consider a full volume fraction.  By considering the phase factors displayed in Table 1, it is clear that such a ground state would have identical magnetic scattering cross-sections at $(002)$ and $(1\bar13)$, which would be proportional to the net polarization of the $\alpha$-chain spins.  An inspection of Fig. 2 will satisfy that this is not the case experimentally.  The reduction of $(002)$ with respect to $(1\bar13)$ implies some polarization of the beta chains along the field, away from their easy-axes.  The cross-section at $(1\bar12)$ is due solely to the component of the $\beta$-chain spins which is AFM ordered and transverse to $\bm Q \rm$, which is reduced from the single-ion $\beta$-chain moment by a factor of $2/3$.  Therefore, even by promoting the minority phase of Cao \it{et al.} \rm to a full volume fraction, we do not recover enough scattering power at $(1\bar12)$ to account for the measured intensity, assuming that $\alpha$ and $\beta$ chain moments have the same magnitude in the ground state.  One would expect the scattering at $(002)$ and $(1\bar13)$ to be dominant in this case.  Worse still, we have the unfortunate problem of requiring a larger than expected AFM moment on the $\beta$-chains while at the same time requiring a finite polarization of these same spins.  Clearly, the classical Ising expectation is not a good approximation to the field-induced ground state realized experimentally.

\begin{table}
\caption{Ordered moment in the unit cell. (0.4K, 7T).}
\label{table:2}
\begin{ruledtabular}
\begin{tabular}{c c c}
 $(002)$ & $(1\bar13)$ & $(1\bar12)$ \\ \hline
 14.0 $\mu_B$ & 20.8 $\mu_B$ & 41.5 $\mu_B$ \\

\end{tabular}
\end{ruledtabular}
\end{table}

We can better characterize the nature of the ordered state by relaxing the Ising constraint.  The necessity of this procedure is not unexpected, since the relatively small anisotropy gap is expected to allow for some restoration of rotational symmetry.\cite{enjalran, hamid, mirebeauice}  In general, the symmetry of the problem suggests that we consider independent $[110]$-field-polarized moments on the $\alpha$ and $\beta$ chain sites, which we will refer to as $\bm{P}_{\alpha}$ and $\bm{P}_{\beta}$ respectively.  In addition, we consider independent AFM ordered moments on each sets of chains: $\bm{S}_{\alpha}$ and $\bm{S}_{\beta}$, which contribute to the scattering at $(1\bar12)$.  In this picture, the scattering at $(002)$ will arise from $|8(P_{\alpha} - P_{\beta})|$, the scattering at $(1\bar13)$ will arise from $|8P_{\alpha}|$, and the scattering at $(1\bar12)$ will probe the  quantity $|8(\bm{S}_{\alpha} + \it{i}$$ \bm{S}_{\beta})|_{\perp}$.  It is important to note that while the polarized moments are transverse to $\bm Q \rm$ since the magnetic field is applied transverse to the scattering plane, for the AFM moments this is not necessarily the case. Only the AFM moment transverse to $\bm Q \rm$ will contribute to the scattering at $(1\bar12)$.  In order to convert the quantities plotted in Fig. 2c,d to Bohr magnetons, we first normalize to the nuclear contribution measured at $(1\bar13)$, for which the scattering length can be calculated for the known crystal lattice in cm$^{-12}$.  This value is then converted to Bohr magnetons using the ratio $(2 / \gamma r_0)$.\cite{billinge}  The resulting total moments transverse to $\bm Q \rm$, summed over the 16 spin unit cell,  are listed in Table 2.  These values constitute lower limits, as extinction effects will act to decrease the measured magnetic moments.  However, since extinction is expected to preferentially suppress strong reflections, one would expect that the dominant effect of a proper extinction correction would be to $\it{increase} \rm$ the AFM ordered moment relative to the polarization.

Now, we are in a position to offer a description of the ordered state.  The net polarization per spin, given by the quantity $(P_{\alpha} + P_{\beta})/2$, can be derived as $\sim$ 1.7 $\mu_B$ if $P_{\alpha} > P_{\beta}$, and $\sim$ 3.5 $\mu_B$ if $P_{\alpha} < P_{\beta}$.  These values are significantly lower than the saturation magnetization reported in the literature,\cite{mag,mirebeauice} which could either indicate that a strong extinction correction is necessary, that precise alignment of the applied magnetic field leads to a reduction in the polarized moment, or both.  As for the AFM ordered moment, if we assume that it is entirely transverse to $\bm Q \rm$, we arrive at an average value of 3.7 $\mu_B$ per spin.  However, this represents an extreme departure from Ising anisotropy.  If we instead assume that the AFM moment remains Ising-like, we arrive at a value of of 4.9 $\mu_B$ per spin.  Implicitly, we have assumed that the moment is equally distributed amongst all spins (i.e. $|\bm{S}_{\alpha}| = |\bm{S}_{\beta}| = \bar{\bm{S}}$).  This is neither necessary or likely.  However, we do so because it is an instructive way to estimate the average value.  If we assume that only the $\beta$-chain spins participate in the AFM order, then we find that $|\bm{S}_{\beta}|$ = 7.8 $\mu_B$.

It bears mentioning that the above discussion of the ordered state, like that of Cao \textit{et al.}\cite{mirebeauice}, is based on unpolarized neutron scattering measurements.  It is therefore possible that the nuclear Bragg scattering could manifest some magnetic field dependence, which would be impossible to deconvolute from the response of the magnetic spins in these measurements.  Tb$_2$Ti$_2$O$_7$ is a known giant magnetostrictive material\cite{aleksandrov}, although no evidence for a magnetic field induced structural phase transition has ever been reported.  Nevertheless magnetoelastic distortion of the lattice has not been ruled out, and may well play a role in the nature of the ordered phase.


In conclusion, we have reported a careful analysis of three neutron Bragg reflections that strongly constrain the character of the ordered phase in Tb$_2$Ti$_2$O$_7$ under the application of magnetic fields along $[110]$.  The dominant scattering originates from $\bm Q \rm \neq 0$ correlations, which are robust in fields as large as 9T, in contrast to the recent report of Cao \textit{et al.}\cite{mirebeauice}  We have developed a general model based on the assumption that this scattering has magnetic origin, and find that the AFM ordered moment is appreciable - conceivably on the order of 4.9 $\mu_B$ per spin if the system retains an Ising-like character.  This represents a roughly threefold increase in the size of the AFM ordered moment relative to that refined in Ref \onlinecite{mirebeauice}.  We contend that this discrepancy arises due to an extreme sensitivity to magnetic field misalignment.  A possible test for this conjecture would be vector-magnet susceptibility measurements of the type reported in Ref \onlinecite{ryuji}.  We hope that this work motivates such further experimental investigation, as well as theoretical work focusing on the nature of the perturbation-induced ordered phases of this enigmatic magnet.

The authors would like to thank Pat Clancy and Michel Gingras for useful discussions and a critical reading of the manuscript, and acknowledge the contributions of John Copley and Yiming Qiu to the measurements performed at DCS.  This work was supported by NSERC of Canada.


%
%






\end{document}